# Elimination of Nonlinear Deviations in Thermal Lattice BGK Models

Yu Chen, Hirotada Ohashi and Mamoru Akiyama

*Department of Quantum Engineering and Systems Science*
*University of Tokyo*
*7-3-1 Hongo, Bunkyo-ku, Tokyo, 113, JAPAN*

## Abstract

We present a new thermal lattice BGK model in D-dimensional space for the numerical calculation of fluid dynamics. This model uses a higher order expansion of equilibrium distribution in Maxwellian type. In the mean time the lattice symmetry is upgraded to ensure the isotropy of 6th order tensor. These manipulations lead to macroscopic equations free from nonlinear deviations. We demonstrate the improvements by conducting classical Chapman-Enskog analysis and by numerical simulation of shear wave flow. The transport coefficients are measured numerically, too.

## 1 INTRODUCTION

Since the proposal of lattice gas automata(LGA) model for the simulation of fluid flow, much attention has been paid to the totally discrete kinetic theory. The governing equation for this relatively new theory is called as the lattice Boltzmann equation(LBE). Although being a derivation of the LGA model, LBE became popular rapidly among the enthusiastic researchers of lattice hydrodynamics because it is noise-free and so flexible that the physics of real fluid could be fully recovered. These advantages may well be observed from the newly suggested lattice BGK model[1], in which the BGK collision operator[2] was introduced to the original LBE models. Today, main reasons that could be forwarded to persuade a fluid dynamic expert to adopt the lattice BGK model as an alternative tool for numerical simulation lie in three folds: the simple algorithm, the easiness in dealing with complicated geometric boundaries and the high level of parallelism in the implementation.

The recent topics about the lattice BGK models concerned with the nonlinear deviation in compressible regime and models that consider thermal effects. The nonlinear deviation was first derived by Qian[3] to be an additional dissipative term in the r. h. s. of the macroscopic momentum equation of the non-thermal lattice BGK model. Hence the macro-dynamics of such model is actually governed by

Navier-Stokes Equation $+ \sigma \partial_\beta \partial_\gamma (\rho u_\alpha u_\beta u_\gamma)$.    (1)

Here $\rho$, $\boldsymbol{u}$ are density and flow velocity. $\sigma$ is defined as $(\frac{1}{2} - \tau)$, where $\tau$ is the relaxation time of the BGK collision operator. When the flow enters into the compressible regime, the deviation becomes significant as the Mach number is no longer small. The pioneer work of the thermal lattice BGK model was published in [4], where the hexongal lattice and 2nd order expansion of equilibrium distribution are employed. The conservation of mass, momentum and energy in micro-world gives rise to macroscopic equations which look like those of real fluid, except for some hidden nonlinear terms.

The current thermal lattice BGK models thus are not free from nonlinear deviation. The deviation in momentum equation is similar to the term appearing in equation (1). Deviations in energy equation are found to be proportional to $\partial_\alpha \partial_\beta (\rho e u_\alpha u_\beta)$, $\partial_\alpha \partial_\beta (\rho u_\alpha u_\beta)$, ..., etc. In order to eliminate these deviations, higher order expansion(4th) has to be considered for the equilibrium distribution function. Furthermore, the lattice should be symmetric enough to ensure isotropy up to 6th order tensor so that higher order expansions could be included correctly. This cannot be realized on lattices consisting of regular polygon(such as hexongal lattice), as they ensure at most the 4th order tensor to be isotropic.

In the sequel, we shall introduce a model which is able to eliminate the nonlinear deviations. We present the theoretical works concerning with the micro- and macro-dynamics in Section 2. In Section 3 the transport coefficients will be numerically measured and the elimination of the nonlinear term in the momentum equation will be illustrated. Section 4 gives the concluding remarks.

## 2 LATTICE HYDRODYNAMICS

### 2.1 Lattice Geometry and Symmetry

As in our previous model[5], the lattice we employed is composed of several sub-lattices. The coordinate of base vectors of these sub-lattices may be represented by $k(\pm 1, \ldots, \pm 1, 0, \ldots, 0)$ and its permutations. The number of non-zero components is recorded as $p$, which also stands for the modulus of such base vector. $k$ is a multiplier for the magnitude of modulus. Hence the vector of lattice link may be denoted as $c_{pki}$, where $pk$ gives the index of sub-lattice and $i$ counts total $b_p$ vectors lie in the $pk$ sub-lattice. The tensor of $n$-velocity moments, defined as $T^{(n)}_{pk\alpha\ldots\xi} = \sum_i c_{pki\alpha}\ldots c_{pki\xi}$, is crucial to the hydrodynamic derivation. The odd order tensors vanish naturally by the definition itself and the 2nd, 4th and 6th order tensors are required to be isotropic. It can be shown that these even order tensors are generally written as

$$T^{(2)}_{pk\alpha\beta} = \frac{b_p c_{pk}^2}{D}\delta_{\alpha\beta},\qquad(2)$$
$$T^{(4)}_{pk\alpha\beta\gamma\delta} = \psi_{pk}\Upsilon_{\alpha\beta\gamma\delta} + \varphi_{pk}(\delta_{\alpha\beta}\delta_{\gamma\delta} + \ldots),$$
$$T^{(6)}_{pk\alpha\beta\gamma\delta\zeta\xi} = \Lambda_{pk}\Upsilon_{\alpha\beta\gamma\delta\zeta\xi} + \Omega_{pk}(\delta_{\alpha\beta}\Upsilon_{\gamma\delta\zeta\xi} + \ldots)$$
$$+ \Theta_{pk}(\delta_{\alpha\beta}T^{(4)}_{pk\gamma\delta\zeta\xi} + \ldots),$$

where $\Upsilon$ and $\delta$ are Kronecker tensors, "..." represents permutations of the indices of the first term. The specific values of $\psi_{pk}$, $\varphi_{pk}$, $\Lambda_{pk}$, $\Omega_{pk}$ and $\Theta_{pk}$ could be calculated by the products of gauge vectors($\pm c_{p1i}$, $\pm c_{p1i} \pm c_{p1j}$, ...) and are listed as follows for two dimensional space,

$$\begin{cases}\psi_{11} = 2 \\ \psi_{21} = -8\end{cases},\begin{cases}\varphi_{11} = 0 \\ \varphi_{21} = 4\end{cases}.\qquad(3)$$

$$\begin{cases}\Lambda_{11} = 2 \\ \Lambda_{21} = -16\end{cases},\begin{cases}\Omega_{11} = 0 \\ \Omega_{21} = 0\end{cases},\begin{cases}\Theta_{11} = 0 \\ \Theta_{21} = \frac{4}{3}\end{cases}.\qquad(4)$$

In case $k \neq 1$, these values should be multiplied by $k^n$. With careful inspection of these parameters, we may infer that anisotropic parts of 4th and 6th order tensors would vanish simultaneously if the ratio of the particle distributions on different sub-lattices is properly tuned.

### 2.2 Lattice BGK Equation

Lattice BGK equation describes the propagation and collision of particles occurring on the discrete spatial lattice, at discrete time step and with discrete velocity set. Taking the particle distribution on each lattice link as $N_{pki}(\boldsymbol{x}, t)$, this equation appears to be

$$N_{pki}(\boldsymbol{x} + \boldsymbol{c}_{pki}, t+1) - N_{pki}(\boldsymbol{x}, t) = \qquad(5)$$
$$-\frac{1}{\tau}(N_{pki} - N^{[eq]}_{pki}).$$

Here the collisions of particles are replaced by a relaxation process, in which the particle distribution is relaxed to its equilibrium value over a time period $\tau$. In order to reproduce Navier-Stokes fluid the equilibrium distribution should be Maxwellian and depends only on the local conserved density of mass, momentum and energy, which are defined by the following formulas,

$$\sum_{pki} N_{pki} = \rho,\ \sum_{pki} N_{pki}c_{pki\alpha} = \rho u_\alpha,\qquad(6)$$
$$\sum_{pki} N_{pki}c_{pki}^2 = \rho(2e + u^2).$$

Here $e$ is the thermal energy determined by flying speed of particles. When the flow velocity is very much smaller than this speed, the equilibrium distribution is expanded into the Chapman-Enskog form with the consideration of parity invariance of regular lattices,

$$\begin{aligned}N^{[eq]}_{pki} &= A_{pk} + M_{pk}(u_\alpha c_{pki\alpha}) + G_{pk}u^2 + \qquad(7)\\ &\quad J_{pk}(u_\alpha c_{pki\alpha})^2 + L_{pk}(u_\alpha c_{pki\alpha})u^2 +\\ &\quad H_{pk}(u_\alpha c_{pki\alpha})^3 + R_{pk}(u_\alpha c_{pki\alpha})^2 u^2 +\\ &\quad S_{pk}u^4 + \mathcal{O}(u^5).\end{aligned}$$

The parameters stated above depend only on local $\rho$ and $e$ and are to be written in this form,

$$X_{pk} = \rho(x_{pk0} + x_{pk1}e + x_{pk2}e^2).\qquad(8)$$

$X_{pk}$ may represent any of $A_{pk}, M_{pk} \ldots$ etc. The specific values of these parameters could be determined by considering higer order isotropy and matching the form of macroscopic equations with those of physical fluids. We summarize constraints to the parameters that would lead to the N-S equation in Table 1.

### 2.3 Macroscopic Equations

Using the constraints listed above and the symmetric properties given in equation (2), we calculated the velocity moments of the equilibrium distribution to 4th order. The obtained results not only verified the definitions of mass, momentum and energy but also gave isotropic momentum and energy fluxes. Our results could be reduced to the ones published in [6], in which only the special three dimensional case was considered. Next we employed the multi-scale technique[5] and derived the macroscopic equations in a straightforward way. The resulting equations are exactly the compressible Navier-Stokes equations(equations(1), (2), (3) in [4]). Here, in D dimensions, we list the equation of state,

$$p = \frac{2}{D}\rho e, \qquad (9)$$

and give expressions for transport coefficients,

$$\begin{aligned}
\mu &= \frac{2}{D}\rho e(\tau - \frac{1}{2}), \qquad (10)\\
\lambda &= -\frac{4}{D^2}\rho e(\tau - \frac{1}{2}),\\
\kappa &= \frac{2(D+2)}{D^2}\rho e(\tau - \frac{1}{2}).
\end{aligned}$$

Here $p$ is the thermodynamic pressure and $\mu$, $\lambda$ and $\kappa$ stand for shear viscosity, second viscosity and thermal conductivity respectively.

We observed from analytical results that terms of nonlinear deviation no longer bother the macrodynamics of the lattice BGK model. Partly because of this and partly because of convenience, we label our model in the sequel as ETLBE(Exact Thermal LBE) and the conventional model TLBE(Thermal LBE).

## 3 NUMERICAL RESULTS

We carried out all the numerical calculation on a two dimensional least ETLBE, in which 16 velocities are employed and indices of sub-lattices are 11, 12, 21 and 22. To show the improvement of the new model, we also coded a 13-velocity TLBE in which equilibrium distribution is expanded to 2nd order and isotropy is kept only to 4th order. The numerical results are compared with each other for every case.

### 3.1 Transport Coefficients

We numerically measured the sound speed, shear viscosity and thermal conductivity by using the linear perturbation theory. Figure 1 is the results of sound speed. The results of two models agree well with the theoretical prediction. The results of measurement of shear viscosity are shown in Figure 2. We observed deviation from theoretical value for both ETLBE and TLBE when $\tau$ gets large enough. This is due to increment of Knudsen number which may break the hydrodynamic mode. It is interesting, however, that the way the deviation increases is different regarding to each model at higher energy level. The heat conductivity is measured by switching off the convection effects. Both the models give good results, see Figure 3.

### 3.2 Elimination of Nonlinear Deviation

We illustrate the elimination by simulating the shear wave flow

$$u_x = u_0, u_y = u_0 e^{i(\omega t + kx)}, \qquad (11)$$

on the 64×64 lattice field. The flow velocity $u_0$ could be changed to adjust Mach number. The Mach number effects on the decaying rate of the total kinetic energy are investigated and shown in Figure 4. We observed that the ETLBE behaves extremely well even $\sqrt{2}Ma$ reaches 0.95. The decaying rates consistently stick to the value of $2\nu k^2$, where $k$ is the wave number. For the TLBE, the deviation is intolerable. However this deviation agrees with the theoretically predicted nonlinear term as well, namely, the total kinetic energy decays as $e^{-2(\nu - \sigma u_0^2)k^2 t}$.

## 4 CONCLUSION

We developed a new thermal lattice BGK model which eliminates the reported nonlinear deviations effectively. This was demonstrated both analytically and numerically. A much detailed comparison between the results of lattice BGK simulation and the Navier-Stokes solution is very much expected.

| | $A_{pk}$ | | | $M_{pk}$ | | $J_{pk}$ | | $G_{pk}$ | | $H_{pk}$ | $L_{pk}$ | $R_{pk}$ | $S_{pk}$ |
|---|---|---|---|---|---|---|---|---|---|---|---|---|---|
| | $a_{pk0}$ | $a_{pk1}$ | $a_{pk2}$ | $m_{pk0}$ | $m_{pk1}$ | $j_{pk0}$ | $j_{pk1}$ | $g_{pk0}$ | $g_{pk1}$ | $h_{pk0}$ | $l_{pk0}$ | $r_{pk0}$ | $s_{pk0}$ |
| $\sum_{pk} b_p x_{pki}$ | 1 | 0 | 0 | | | | | $-\frac{X_0}{D}$ | $-\frac{X_1}{D}$ | | | | $-\frac{X_3}{D}$ |
| $\sum_{pk} b_p c_{pk}^2 x_{pki}$ | 0 | 2 | 0 | $D$ | 0 | $X_0$ | $X_1$ | $-\frac{D}{2}$ | 0 | | $-3DX_2$ | $X_3$ | 0 |
| $\sum_{pk} \psi_{pk} x_{pki}$ | 0 | 0 | 0 | 0 | 0 | 0 | 0 | 0 | 0 | 0 | 0 | 0 | 0 |
| $\sum_{pk} \varphi_{pk} x_{pki}$ | 0 | 0 | $\frac{4}{D^2}$ | 0 | $\frac{2}{D}$ | $\frac{1}{2}$ | 0 | 0 | $-\frac{1}{D}$ | $X_2$ | $-\frac{1}{2}$ | 0 | $-\frac{1}{2(D+2)}$ |
| $\sum_{pk} \Lambda_{pk} x_{pki}$ | | | | 0 | 0 | | | | | 0 | | 0 | |
| $\sum_{pk} \Omega_{pk} x_{pki}$ | | | | 0 | 0 | | | | | 0 | | 0 | |
| $\sum_{pk} \Theta_{pk} x_{pki}$ | | | | 0 | $\frac{1}{D}$ | | | | | $\frac{1}{6}$ | | $\frac{1}{2(D+4)}$ | |

Table 1: Constraints to the parameters of expanded $N_{pki}^{[eq]}$. Blanks mean no constraints. $X_0...X_3$ are not constraints, they are used later to calculate other parameters.

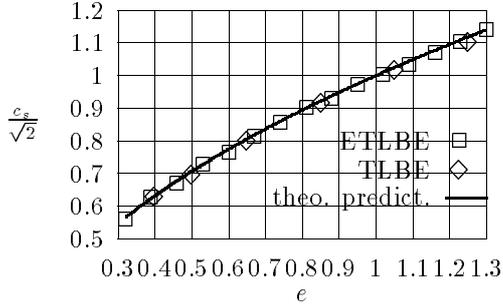

Figure 1: Sound speed as a function of internal energy $e$.

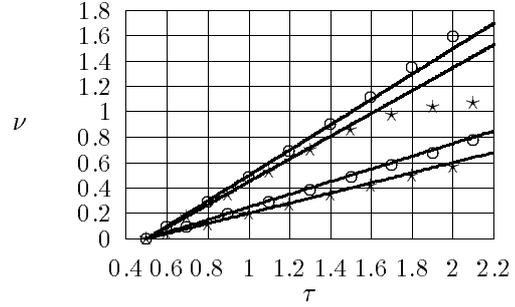

Figure 2: Shear viscosity as a function of relaxation time $\tau$. $e = 0.4, 0.5, 0.9, 1.0$, star: TLBE; circle: ETLBE; line: Theory.

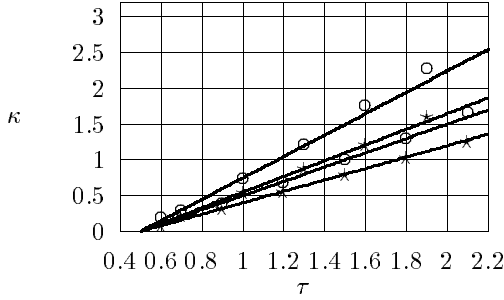

Figure 3: Thermal conductivity as a function of relaxation time $\tau$. $e = 0.4, 0.5, 0.55, 0.75$, star: TLBE; circle: ETLBE; line: Theory.

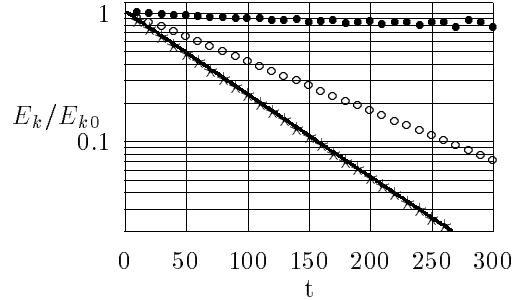

Figure 4: $Ma$ effects on the decaying rate of total kinetic energy in shear wave flow. $\sqrt{2}Ma = 0.0158, 0.63, 0.95$, dots: TLBE; lines: ETLBE.